# Direct Multispectral Satellite Nowcasting of Gridded Surface Solar Radiation

Tianyu Xie[1,2#], Zhaoliang Zeng[1,2#]*, Zhehao Liang[3], Ke Gui[1]*,Xutao Zhang[1], Yaqiang Wang[1,2], Dabo Guan[4,5], Huizheng Che[1]

[1] State Key Laboratory of Severe Weather Meteorological Science and Technology, Chinese Academy of Meteorological Sciences, Beijing, China

[2] Xiong'an Institute of Meteorological Artificial Intelligence, Xiong'an, Hebei, China

[3] Department of Applied and Computational Mathematics and Statistics, University of Notre Dame, Notre Dame, IN 46556, USA

[4] Department of Earth System Sciences, Tsinghua University, Beijing 10084, China

[5] The Bartlett School of Sustainable Construction, University College London, London WC1E 7HB, U.K.

# These authors contributed equally.

* Corresponding authors.

E-mail Addresses: zengzl@cma.gov.cn (Z. Zeng) and guik@cma.gov.cn (K. Gui)

## Abstract

Surface shortwave radiation (SWR) can change within minutes as clouds evolve, but operational forecasts often trade spatial detail and update frequency for forecast range. Here, we present a numerical-weather-prediction-independent, observation-to-forecast framework that converts sequences of Himawari-9 Level-1 multispectral imagery and solar geometry directly into 24 gridded SWR fields at 10-min intervals for the next 4 h. A controlled six-model experiment in eastern China separates the effects of satellite information from those of temporal self-attention. In the 0-1 h window, the full-multispectral attention model achieves a mean absolute error of 37.9 W/m$^2$ and a root mean square error of 63.7 W/m$^2$, reductions of 21.0% and 19.5% relative to a historical-SWR baseline. Improvements extend across nine subregions and include weaker spatial bias and better structural consistency. Temporal attention does not improve forecasts from historical SWR alone, but becomes effective when supplied with multispectral cloud and thermodynamic signals. Direct multispectral observations therefore provide the precursor information required for high-frequency, spatially resolved SWR nowcasting.

## Introduction

Surface shortwave radiation (SWR) regulates the surface energy budget and influences land-atmosphere exchange, ecosystem productivity and solar-power generation[1–3]. The rapid expansion of photovoltaic capacity has increased the need for radiation forecasts that resolve cloud-driven changes at the spatial and temporal scales relevant to grid operations[4–6]. Numerical weather prediction (NWP) remains essential beyond the nowcasting range, but its model cycles and typical output grids can be too coarse to capture cloud evolution that changes surface radiation within minutes[4,5,7,8]. This limitation is most consequential near cloud edges and during cloud formation or dissipation, when SWR can exhibit sharp spatial gradients and rapid temporal ramps. Although aerosol loading can also influence the relationship between satellite-observed radiances and surface SWR, its effects are not explicitly isolated in the present study, which instead focuses on the short-term, cloud-driven evolution of surface radiation[9–15]. In this context, eastern China is selected as the study region for evaluating high-frequency, spatially resolved SWR nowcasting.

Geostationary satellites provide a direct and frequently updated view of the cloud and thermodynamic states that control these changes. Himawari-9/AHI observes East Asia in 16 spectral channels at high cadence[16], and its reflectance, brightness-temperature and solar-geometry information supports high-resolution SWR retrieval[17,18]. Satellite-based forecasts have subsequently extrapolated retrieved irradiance or cloud-index fields, learned future cloud or satellite images, and predicted irradiance for individual sites or local patches[19–21]. These studies establish the value of satellite observations, but leave a methodological gap. Forecasting a compressed cloud or irradiance proxy can discard complementary spectral information, whereas site-scale predictions do not provide a regional radiation field. It therefore remains unclear how much predictive information is retained when Level-1 multispectral sequences are mapped directly to multi-step, gridded SWR forecasts without NWP or an explicit cloud-motion stage. More broadly, deep learning has been identified as a promising approach for capturing spatial and temporal context in Earth-system observations, including dependencies that may be missed by single-variable or hand-crafted representations[22].

Here we develop an observation-to-forecast framework that maps 24 historical Himawari-9 frames to 24 future SWR fields over eastern China at 10-min resolution. The inputs are evaluated at three levels: historical SWR alone, ten selected Level-1 and derived variables, and the complete Level-1 multispectral set with solar geometry. Each input is tested with and without temporal self-attention in the same SimVP-Swin backbone. This factorial design places the direct multispectral forecasting pathway at the centre of the study while using the sensitivity experiments to test whether skill arises from richer observations, the attention mechanism or their interaction. The results show that multispectral observations improve error, spatial structure and bias across the domain, and that temporal attention is beneficial only when the input contains cloud-radiation evolution signals.

## Results

We evaluate six configurations on the independent test set. Exp1 uses historical SWR only; Exp2 uses ten selected Himawari-9 Level-1 and derived variables; and Exp3 uses the complete Level-1 set of six albedo channels, ten brightness-temperature channels, solar zenith angle and solar azimuth angle. For each input, noTA denotes the SimVP-Swin backbone and TA denotes the same backbone with temporal self-attention. All configurations use the same sample construction and evaluation protocol described in Methods.

### Forecast performance across lead times

Across most lead times, Exp2-TA and Exp3-TA have lower MAE and RMSE and higher R and SSIM than the other configurations (**Fig. 1**). Exp1, which receives only historical SWR, has larger errors and stronger lead-time fluctuations. Multispectral inputs improve the noTA backbone, but the corresponding TA configurations generally perform better. The ranking shows that the attention gain is conditional on the information supplied to the model rather than being an input-independent architectural effect.

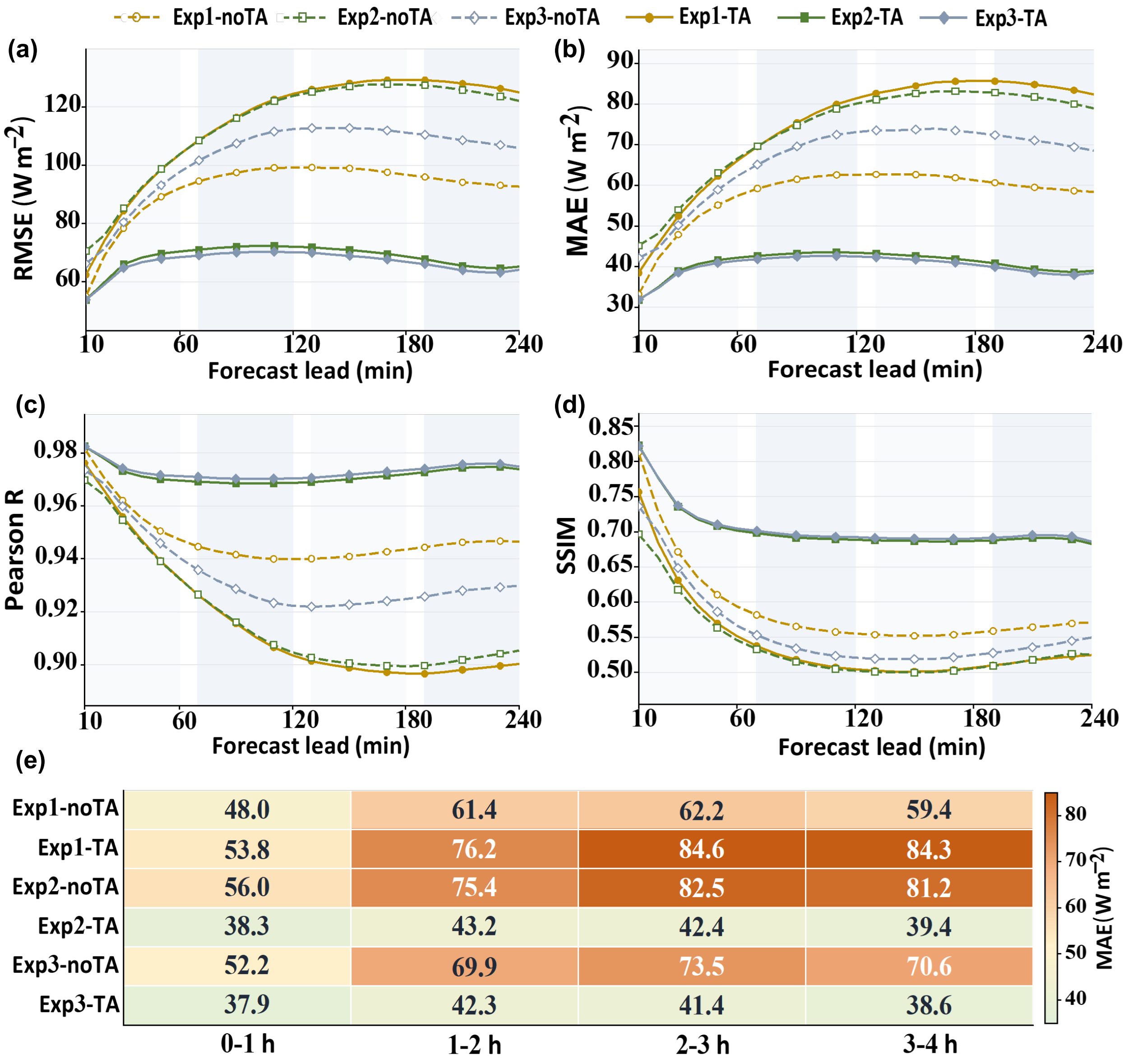

**Fig. 1 | Forecast performance on the independent test set. a-d,** Root mean square error (RMSE), mean absolute error (MAE), Pearson correlation coefficient (R) and structural similarity index measure (SSIM) for the six configurations at 24 lead times. Shading marks the 0-1 h, 1-2 h, 2-3 h and 3-4 h windows. **e,** MAE summarized for the four windows.

In the 0-1 h window, Exp3-TA achieves the best overall performance, with an MAE of 37.9 W/m$^2$, an RMSE of 63.7 W/m$^2$, an R of approximately 0.975 and an SSIM of approximately 0.745. Relative to Exp1-noTA, its MAE and RMSE are lower by 21.0% and 19.5%, respectively. Exp2-TA reaches an MAE of 38.3 W/m$^2$ and an RMSE of 64.9 W/m$^2$, indicating that the selected variables retain most of the predictive information in the full Level-1 set. Its MAE in the 1-2 h, 2-3 h and 3-4 h windows is approximately 43.2, 42.4 and 39.4 W/m$^2$; Exp3-TA is similar and marginally better for most metrics. The apparent decrease in later-window error reflects the diurnal distribution of verification times and should not be interpreted as increasing skill with lead time.

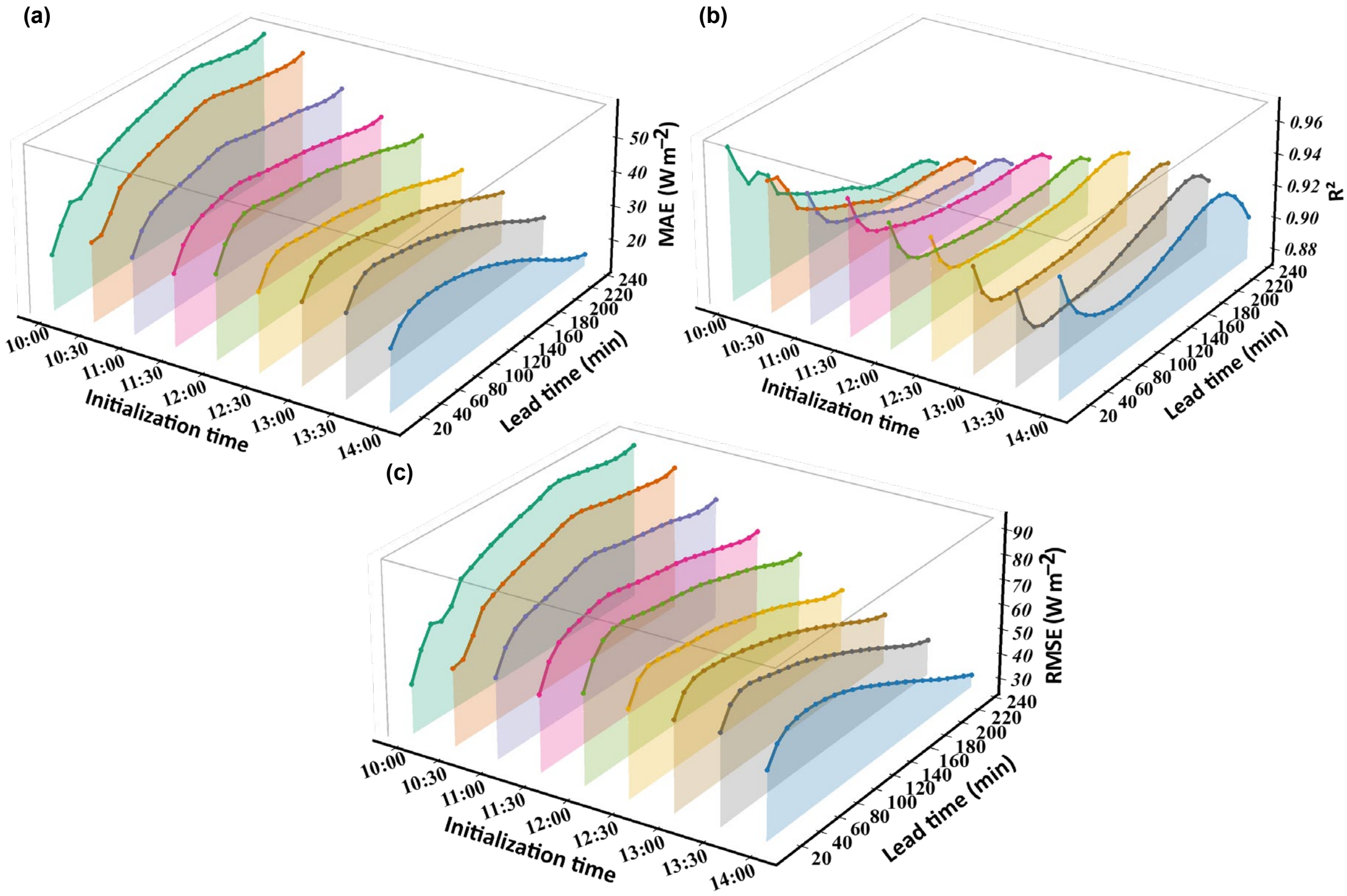

**Fig. 2 | Dependence of Exp3-TA performance on initialization and lead time. a-c,** RMSE, coefficient of determination ($R^2$) and MAE as functions of initialization time and lead time. Lead times extend from 10 to 240 min at 10-min intervals; each coloured curve represents one initialization-time group.

When all samples are pooled, RMSE and MAE are 61.41 and 35.53 W/m$^2$ in the 3-4 h window, compared with 68.63 and 40.70 W/m$^2$ in the 0-1 h window. We therefore group samples by initialization time and follow each group across lead times rather than interpreting the pooled window averages in isolation. **Fig. 2** separates lead-time effects from changes in the diurnal verification background.

The initialization groups show a consistent diurnal structure. For forecasts initialized at 10:00, RMSE and MAE change only slightly, from 89.74 and 56.14 W/m$^2$ in the first window to 88.11 and 55.12 W/m$^2$ in the last. For the 14:00 group, they decrease from 48.52 and 27.63 W/m$^2$ to 35.19 and 17.90 W/m$^2$. Error maxima also shift to shorter lead times for later initializations: RMSE peaks at 160, 140, 110, 80 and 50 min for the 10:00, 10:30, 11:00, 11:30 and 12:00 groups, respectively, but these peaks correspond to a narrow verification period around 12:40-12:50. Thus, SWR error is strongly modulated by verification time, and operational assessment should report initialization time, lead time and verification time together.

## Regional error and spatial bias

We divide the domain into nine subregions (NW, NC, NE, CW, CC, CE, SW, SC and SE) and calculate MAE in each forecast window (Fig. 3). This analysis tests whether the domain-mean improvement is spatially consistent and identifies regions in which forecast error remains systematically high[23].

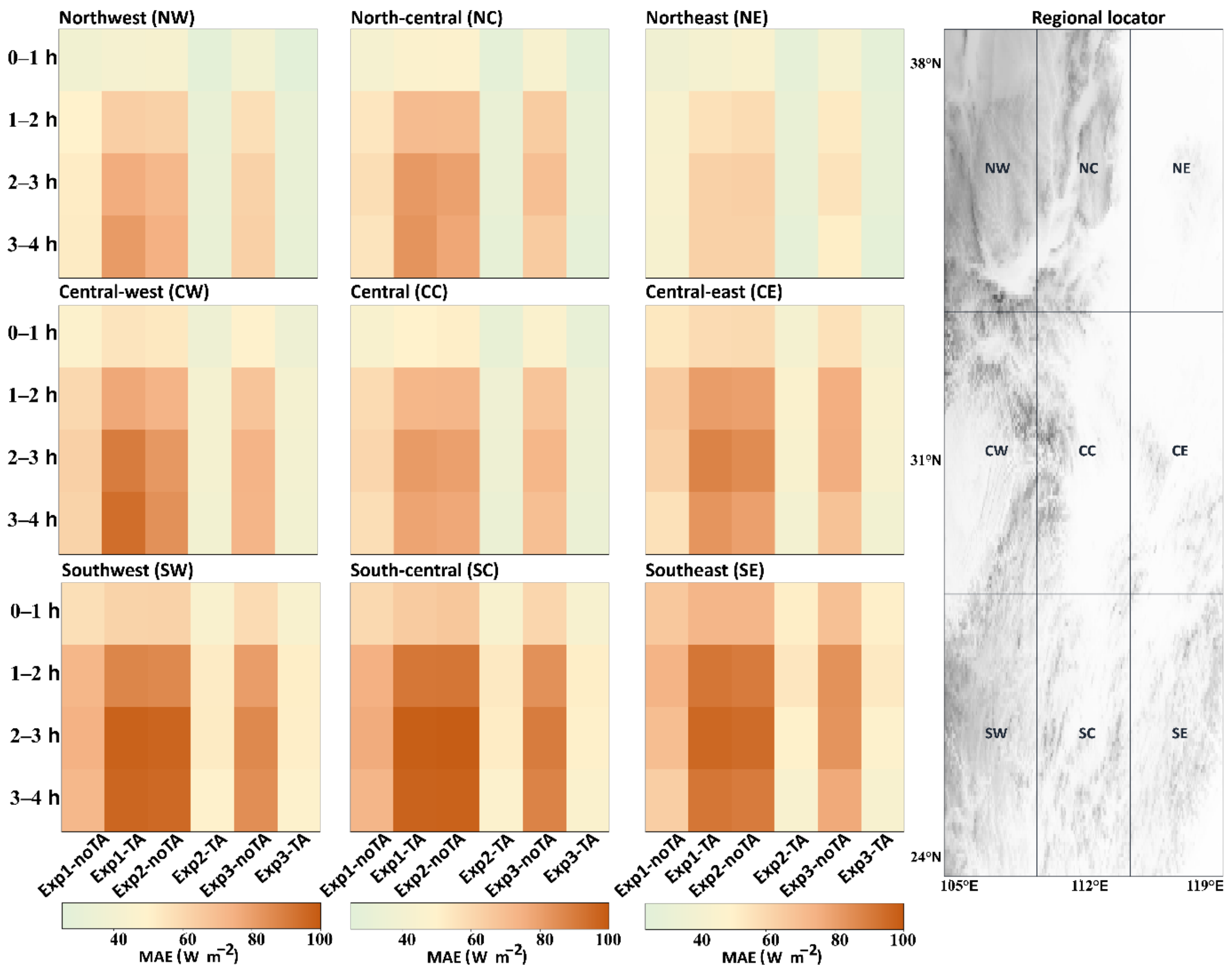

**Fig. 3 | Regional MAE for nine subregions.** Columns show the six configurations and rows show the four forecast windows. Each subregion is represented by a 4 × 6 heat map. On the right is a map of the terrain of nine regions drawn based on elevation files, with shades of gray representing elevation magnitude.

All configurations show substantial spatial heterogeneity. Mean MAE is higher in the three southern regions than in the central and northern regions, and this contrast persists across architectures. The pattern therefore cannot be attributed to one model configuration alone and probably reflects regional differences in cloud variability and SWR predictability. Exp1-noTA, Exp1-TA, Exp2-noTA and Exp3-noTA have higher MAE in most regions, especially in the 2-4 h windows. The regional mean MAE of Exp1-TA is approximately 75.0 W/m$^2$, confirming that temporal attention does not help when historical SWR is the only input. Exp2-TA and Exp3-TA reduce the regional mean to approximately 40.8 and 40.1 W/m$^2$, respectively, and improve all nine subregions. Errors remain largest in the south, but the multispectral TA configurations reduce both their magnitude and their growth across forecast windows.

The regional results demonstrate that the gain from combining Level-1 multispectral inputs with temporal attention is distributed across the domain rather than concentrated in a few locations. The small difference between Exp2-TA and Exp3-TA further indicates that a compact set of physically informative channels captures most of the available signal.

The configurations differ in both bias magnitude and spatial organization (**Fig. 4**). All noTA models contain coherent positive and negative structures, showing that their errors are not spatially random. Exp1-noTA and Exp3-noTA, in particular, produce persistent bands of overestimation or underestimation in several regions and forecast windows. Bias is strongest for Exp1-TA in the middle and later windows, consistent with its high MAE. Temporal attention alone therefore neither reduces absolute error nor removes systematic bias from SWR-only inputs. In contrast, Exp2-TA and Exp3-TA remain closer to zero and contain fewer large, spatially continuous bias regions. Exp3-TA is generally the smoothest and least biased configuration, although adjacent positive and negative patches remain near small-scale gradients.

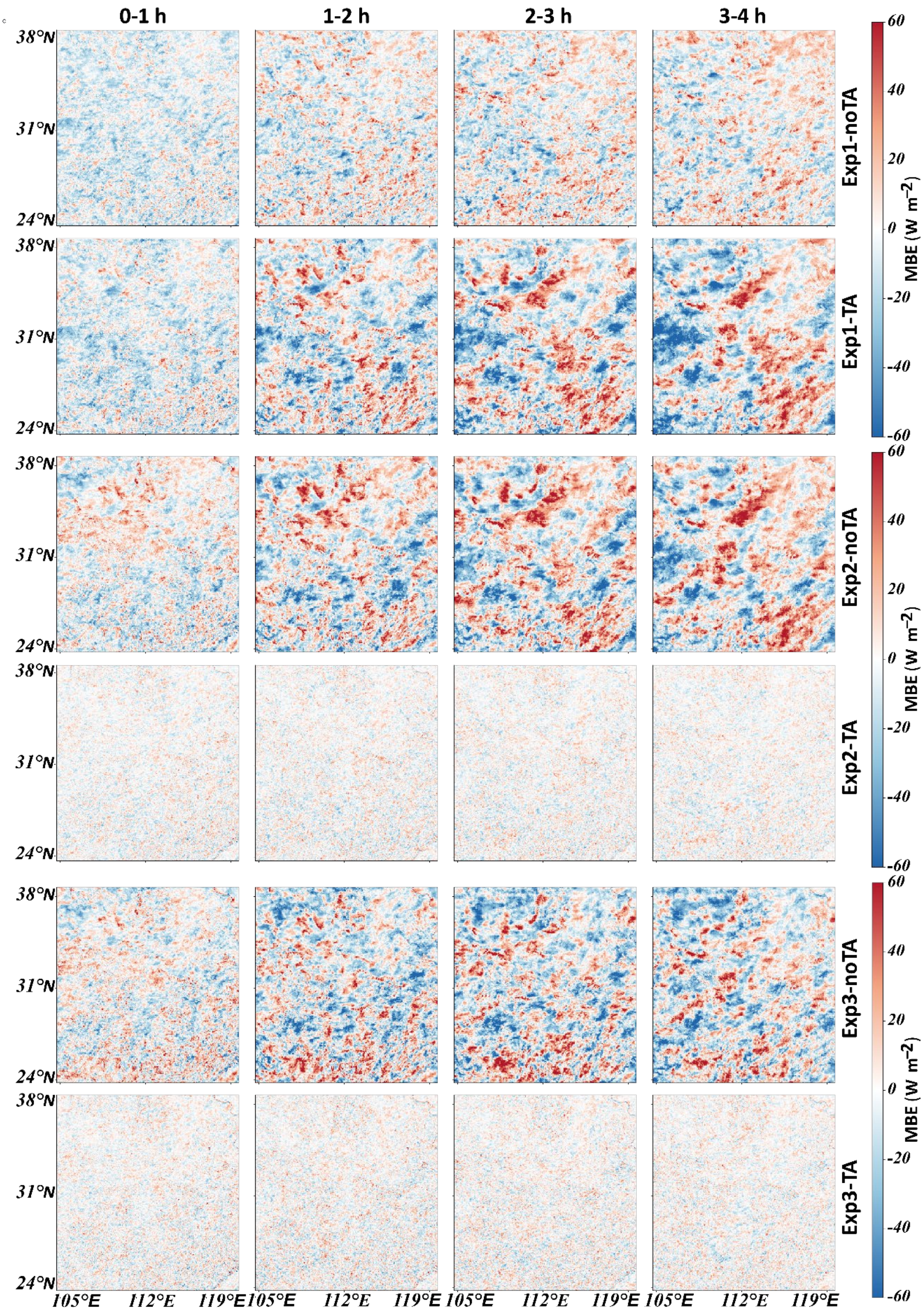

**Fig. 4 | Spatial mean bias error for the six configurations.** Positive values indicate overestimation relative to the Himawari-9 Level-2 SWR target and negative values indicate underestimation.

## Case study of cloud driven-forecast errors

The 2 January 2024 case illustrates how Exp3-TA represents a rapidly evolving radiation field (**Fig. 5**). The model reproduces the large-scale distribution and principal gradients of the Level-2 target, and the main high- and low-SWR regions remain aligned through the forecast. Errors cluster near cloud edges and in cloud-affected areas, where the prediction becomes smoother and local contrast weakens. The model therefore preserves synoptic and mesoscale structure more reliably than small cloud elements and sharp radiation gradients. **Supplementary Fig. 2** and **Supplementary Fig. 3** show the corresponding Exp1-TA and Exp2-TA forecasts. Exp1-TA loses local contrast more rapidly and becomes increasingly smooth at longer lead times. Exp2-TA retains the main radiation pattern but resolves cloud edges and small-scale gradients less effectively than Exp3-TA.

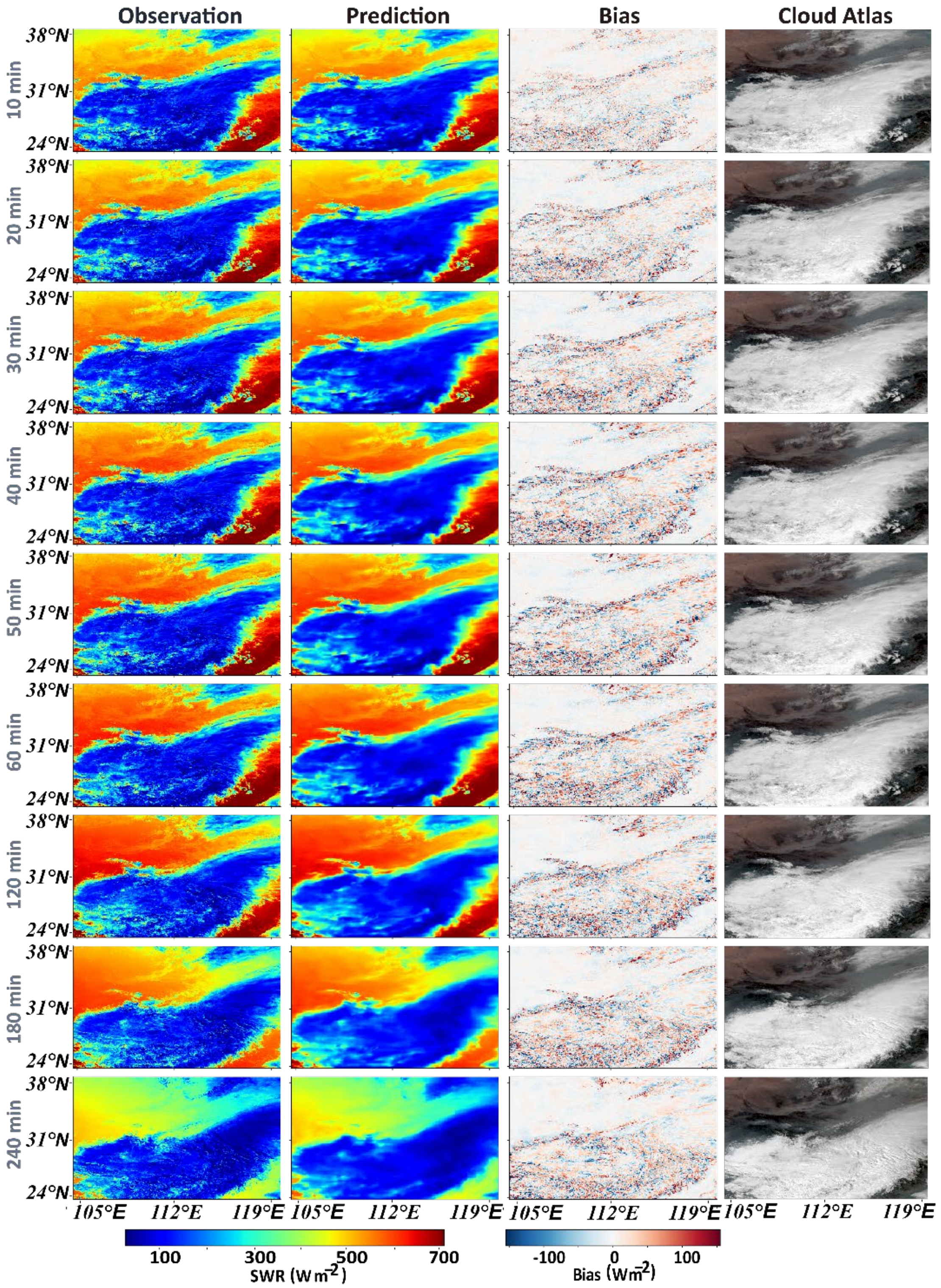

**Fig. 5 | Exp3-TA forecast for 2 January 2024.** Columns show the Himawari-9 Level-2 SWR target, model forecast, forecast-minus-target difference and corresponding cloud image. Rows show selected lead times.

# Discussion

The central advance is a direct observation-to-forecast pathway from high-cadence Level-1 multispectral images to a sequence of regional SWR grids. Earlier satellite nowcasts have forecast retrieved irradiance or cloud proxies, generated future cloud imagery before irradiance conversion, or produced local and site-scale predictions[19–21,24–26]. Here one multi-horizon model produces the complete 10-240 min SWR sequence at 10-min increments without NWP input, explicit cloud-motion vectors or recursive proxy-field extrapolation. Using Level-1 observations as the predictive state preserves spectral signatures of cloud optical and thermodynamic evolution that are unavailable in a single historical SWR field.

The controlled experiments explain why direct multispectral inputs improve the forecast. For Exp1, ΔMAE is positive in multiple regions and windows, so temporal self-attention does not reliably improve an input that contains only past radiation responses (**Fig. 6**). For Exp2 and Exp3, ΔMAE is predominantly negative. Reflectance and brightness-temperature channels expose cloud optical and thermodynamic states before their full radiative effect appears in the target field, and the attention module can assign different weights to these evolving states. The resulting gain is therefore an interaction between observational content and temporal modelling, not a generic consequence of adding model capacity.

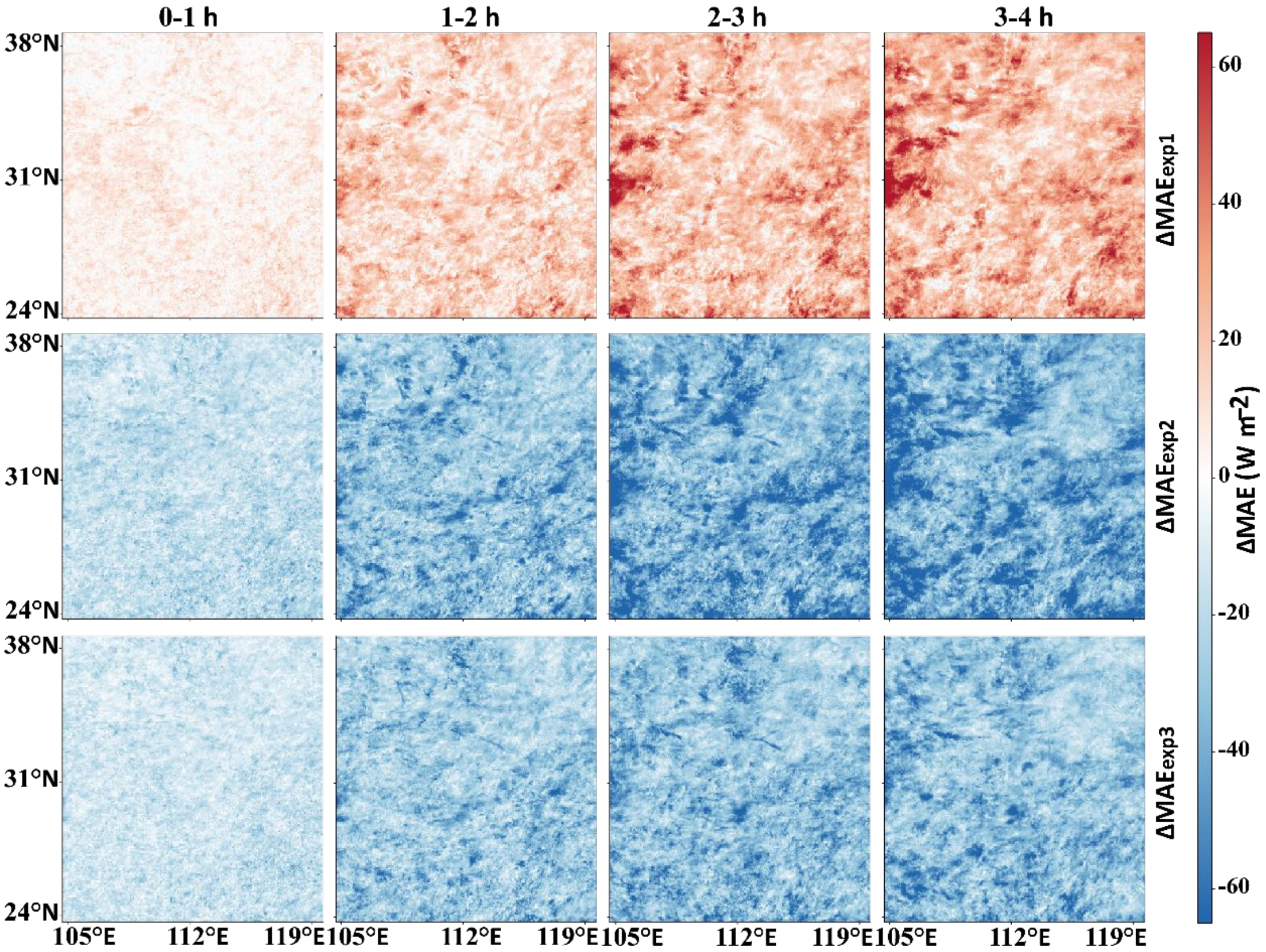

**Fig. 6 | Effect of temporal self-attention for each input experiment.** ΔMAE = MAE_TA − MAE_noTA; negative values indicate that temporal self-attention reduces error.

The similar performance of Exp2-TA and Exp3-TA also constrains the information requirement. Albedo channels encode shortwave reflection by clouds and the surface, brightness temperatures describe cloud-top and clear-sky thermal states, and solar geometry supplies the changing illumination background[27,28]. The selected ten-variable input retains most of the gain from the full Level-1 set, indicating that complementary spectral and geometric information matters more than channel count. This result supports physically informed input compression for regional nowcasting, which can reduce data movement and computation without discarding the main predictive signal.

The input statistics support this interpretation (**Fig. 7** and **Supplementary Fig. 1**). Albedo_03 is negatively correlated with SWR (R approximately −0.448; 7.7% of the summed absolute correlations), whereas tbb_11 is positively correlated (R approximately 0.444; 7.6%). Comparable contributions from several channels show that no single measurement dominates the observed radiation variability[29–31]. These coefficients describe marginal linear association and are not causal effects or model feature importance. The weak raw correlation of solar zenith angle, despite the high selection score of cos (solar zenith angle), further shows that raw correlation, transformed representation and predictive utility are distinct.

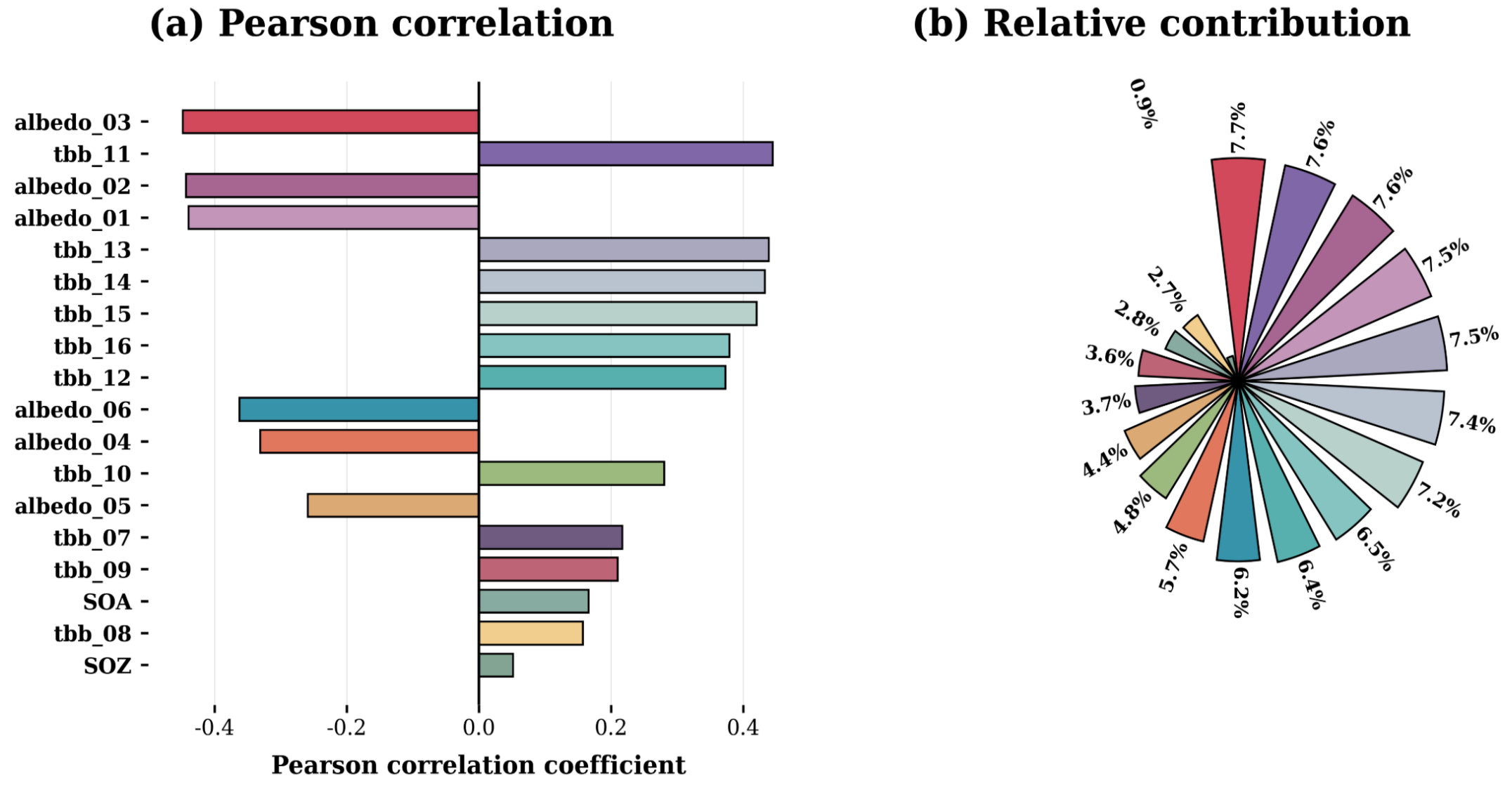

**Fig. 7 | Pearson correlations and relative contributions of 18 variables to SWR. a,** Pearson correlation coefficients between SWR and the candidate variables, ranked from largest to smallest by the absolute coefficient. Bars extending to the right indicate positive correlations and bars extending to the left indicate negative correlations. **b,** Relative contributions of the variables to SWR correlation, calculated by normalizing the absolute Pearson coefficients.

Several limitations define the present scope. Predictions become smooth around small cloud elements, cloud edges and strong radiation gradients, probably because a finite input sequence cannot fully resolve rapid cloud development and because mean-squared-error training penalizes displaced extremes. Similar loss of small-scale spatial detail has been documented in deep-learning-based radar precipitation nowcasting, where probabilistic generative modelling has been shown to produce sharper and more spatially coherent forecasts than conventional deterministic approaches[32]. Although precipitation and SWR involve different physical processes, this finding motivates future investigations of probabilistic or structure-aware training objectives for preserving cloud-edge gradients in SWR forecasts.

The target is the Himawari-9 Level-2 SWR retrieval, so the reported errors quantify forecast agreement with that product rather than independent surface truth. Validation against radiometers or photovoltaic-plant measurements is required before operational use[33,34]. Tests in other climate zones, seasons and terrain regimes are also needed[35–37]. Within these limits, the study establishes a direct path from high-cadence multispectral observations to multi-hour gridded radiation forecasts and shows that its skill depends on observable cloud-radiation precursors.

# Methods

## Data and forecast sample construction

We construct SWR nowcasting samples from Himawari-9 observations acquired during 2024. Inputs are Level-1 multispectral variables, and the prediction target is the Himawari-9 Level-2 SWR retrieval. The eastern China domain is resampled to 0.05° and cropped to a regular 300 × 300 grid (**Fig. 8a**). Only observations between 06:00 and 18:00 Beijing Time are retained to exclude nighttime scenes. We apply physical-range screening, clipping where required, and a common mask for missing or invalid pixels. Input variables are standardized with training-set means and standard deviations; the SWR target uses a separate training-set scaler and is transformed back to $W/m^2$ for evaluation. Each sample contains 24 historical frames and 24 target frames at 10-min intervals, giving a 4-h input sequence and lead times of 10-240 min. For every month, days 1-3 form the independent test set, days 4-6 form the validation set and the remaining days form the training set. Figure 8b summarizes the sample timeline and input experiments.

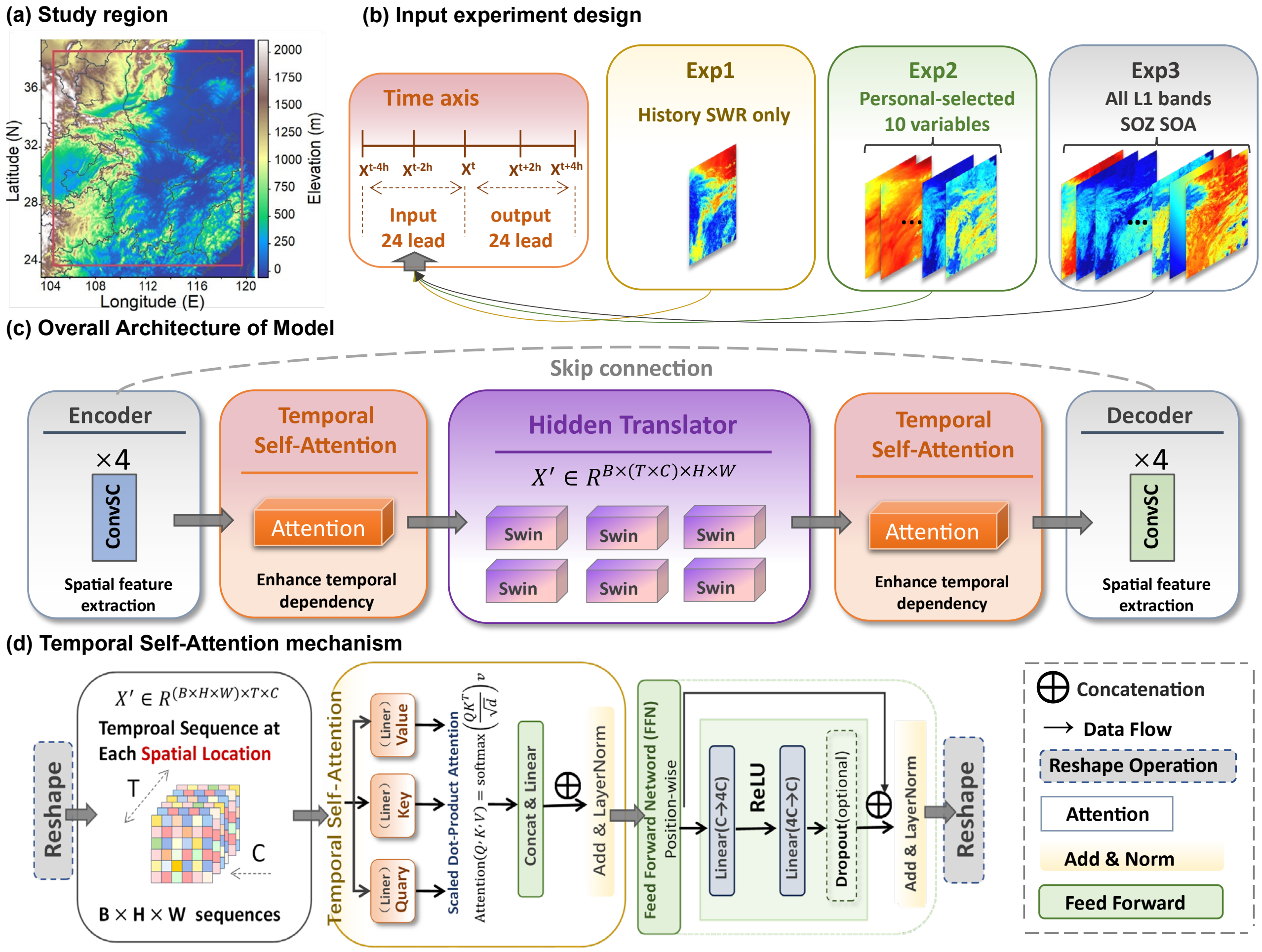

**Fig. 8 | Observation-to-forecast workflow. a,** Study domain. **b,** Input configurations and sample timeline. **c,** SimVP-Swin backbone. **d,** Temporal self-attention module.

Candidate inputs include raw reflectance, brightness temperature and solar-geometry variables, together with eight derived variables: albedo ratios, normalized albedo contrast, cosine of solar zenith angle, relative azimuth angle, brightness-temperature differences and albedo-geometry interaction terms. These variables describe spectral contrast and illumination geometry and are not assigned a causal interpretation. We rank candidates with a combined score based on Pearson correlation, Spearman correlation, Kendall's tau and mutual information:

$$S = 0.4|r_P| + 0.3|r_S| + 0.2|\tau| + 0.1\widetilde{MI} \quad (1)$$

where $\widetilde{MI}$ denotes mutual information normalized within the candidate-variable set. The combined score measures linear, rank-based and nonlinear dependence between each candidate variable and SWR; pairwise correlations among candidate variables are also examined to identify redundancy[21,38–40]. The ten selected input features are SOZ, cos (SOZ), albedo_01, albedo_02, albedo_04, albedo_05,

tbb_07, tbb_14, tbb_15 and tbb_diff_14_15. The selection results and redundancy structure of the candidate variables are shown in **Supplementary Fig. 1** .

**SimVP-Swin model with temporal attention**

We use SimVP-Swin as the backbone spatiotemporal forecasting model and introduce Temporal Self-Attention into its latent-space temporal modelling component. The model learns a nonlinear mapping from historical multichannel observations to future SWR sequences. For any input configuration, the model input is denoted by $X \in \mathbb{R}^{T_{\mathrm{in}} \times C \times H \times W}$, where $T_{\mathrm{in}} = 24$ is the number of input time steps, C is the number of input channels, and H and W are the spatial dimensions; the model output is denoted by $\hat{Y} \in \mathbb{R}^{T_{\mathrm{out}} \times 1 \times H \times W}$, where $T_{\mathrm{out}} = 24$ and the output corresponds to 24 future SWR forecast fields. The overall architecture is shown in **Fig. 8c**. SimVP-Swin adopts an encoder–hidden-translator–decoder structure. The encoder spatially encodes the input field at each historical time step and uses convolutional spatial blocks to map the raw observations into a latent feature space, extracting local textures and spatial structures from cloud fields, radiation fields and multispectral variables. The hidden translator then models the spatiotemporal evolution of the encoded features in latent space and uses the local-window attention and shifted-window mechanism of the Swin Transformer for spatial feature interaction, thereby representing cloud edges, radiation-field gradients and local spatial variability[41,42].

The baseline SimVP-Swin mixes information across time implicitly in latent space. The TA configuration adds temporal self-attention before and after the hidden translator. At each spatial location, latent features from the historical frames form temporal tokens. Multi-head self-attention relates these tokens, after which residual connections, normalization and feed-forward layers update the representation. The operation explicitly models temporal dependence without discarding the encoded spatial structure (**Fig. 8d**)[43].

Combining the three inputs with the two architectures produces Exp1-noTA, Exp1-TA, Exp2-noTA, Exp2-TA, Exp3-noTA and Exp3-TA. Exp1 tests extrapolation from historical SWR. Exp2 tests whether a compact set of multispectral and solar-geometry variables supplies additional predictive information. Exp3 uses all Level-1 channels and both solar angles. The decoder reconstructs 24 future SWR fields from the latent features. Training minimizes mean squared error in standardized SWR space; evaluation transforms predictions to W/m$^2$ and compares them with the Level-2 targets.

**Evaluation metrics and diagnostic analyses**

Model performance is evaluated on the independent test set. For any model configuration, let y-hat and y denote the prediction and the corresponding L2 SWR target, respectively. The error is defined as $e_i = \hat{y}_i - y_i$, where $i$ indexes a valid sample–pixel pair. For each forecast lead time, all valid test samples and valid pixels are included in the statistics. We use mean absolute error (MAE), root mean square error (RMSE), mean bias error (MBE) and the Pearson correlation coefficient (R) to evaluate the predictions, quantifying error magnitude, the influence of larger errors, systematic over- or underestimation, and linear agreement between predictions and targets, respectively. MAE, RMSE and MBE are reported in W/m$^{-2}$; MBE > 0 indicates overall overestimation and MBE < 0 indicates overall underestimation. All metrics are calculated over corresponding valid sample–pixel pairs[44,45].

Spatial structure is evaluated with the structural similarity index measure (SSIM). For each sample and lead time, SSIM is calculated over local windows as follows:

$$SSIM(x,y) = \frac{(2\mu_x\mu_y + C_1)(2\sigma_{xy} + C_2)}{(\mu_x^2 + \mu_y^2 + C_1)(\sigma_x^2 + \sigma_y^2 + C_2)} \tag{2}$$

where $x$ and $y$ denote the local windows of the forecast and target fields,; $\mu_x$ and $\mu_y$ are local means; $\sigma_x^2$ and $\sigma_y^2$ are local variances; $\sigma_{xy}$ is the local covariance; and $C_1$ and $C_2$ are stabilizing constants. We use a $7 \times 7$ local window and a fixed data range of 1300 W/m$^{-2}$ for SSIM. A window contributes to the mean only when all pixels within it are valid. The final SSIM is the mean over all valid sample–lead fields[46].

Forecast lead times extend from 10 to 240 min at 10-min intervals. For each forecast window, MAE, RMSE, MBE and Pearson $R$ are recalculated after pooling all valid sample–pixel pairs within that window, rather than averaged from lead-specific metrics. SSIM is first calculated for each two-dimensional sample–lead field and then averaged over all valid fields in the window.

To analyse the spatial heterogeneity of error, the study domain is further divided into nine subregions. For region $A$ and forecast lead $t$, regional MAE is defined as:

$$MAE_{A,t} = \frac{1}{N_{A,t}} \sum_{i \in A,t} |\hat{y}_i - y_i| \tag{3}$$

where $N_{A,t}$ is the number of valid sample–pixel pairs in region $A$ at lead $t$. This metric is used to compare forecast-error magnitude among spatial regions.

For the spatial error maps, the absolute errors of all valid samples and leads within a specified forecast window are aggregated at each grid point. For grid point $g$ and forecast window $W$, spatial MAE is defined as:

$$\mathrm{MAE}_{g,W} = \frac{\sum_{s} \sum_{t \in W} M_{s,g,t} \mid \hat{y}_{s,g,t} - y_{s,g,t} \mid}{\sum_{s} \sum_{t \in W} M_{s,g,t}} \quad (4)$$

where $s$ denotes a test sample and $M_{s,g,t}$ is the validity mask. $M_{s,g,t} = 1$ when both the prediction and target are valid and 0 otherwise. This calculation directly aggregates errors over all valid sample–lead pairs and avoids additional weighting caused by different numbers of valid samples across lead times or grid points.

The spatial error difference between models is denoted by $\Delta\mathrm{MAE}$. For models $m_1$ and $m_2$, the difference at grid point $g$ and forecast window $W$ is defined as:

$$\Delta\mathrm{MAE}_{g,W} = \mathrm{MAE}_{g,W}^{m_1} - \mathrm{MAE}_{g,W}^{m_2} \quad (5)$$

When $\Delta\mathrm{MAE} < 0$, the former model has lower error than the latter; when $\Delta\mathrm{MAE} > 0$, the former model has higher error. Spatial-bias analysis calculates MBE at grid-point or regional scale to identify the distribution of systematic over- or underestimation.

## Data availability

Himawari-9 Level-1 and Level-2 data were obtained from the Japan Aerospace Exploration Agency (JAXA) Himawari Monitor P-Tree System (https://www.eorc.jaxa.jp/ptree/).

## Acknowledgements

This research was supported by Beijing Natural Science Foundation (L2609040), the National Natural Science Foundation of China (42675225), and the Special Project for Science and Technology Innovation of Xiong'an New Area (2024XAGG0007).

## Competing interests

The authors declare that they have no competing interests.

# ***Supplementary information***

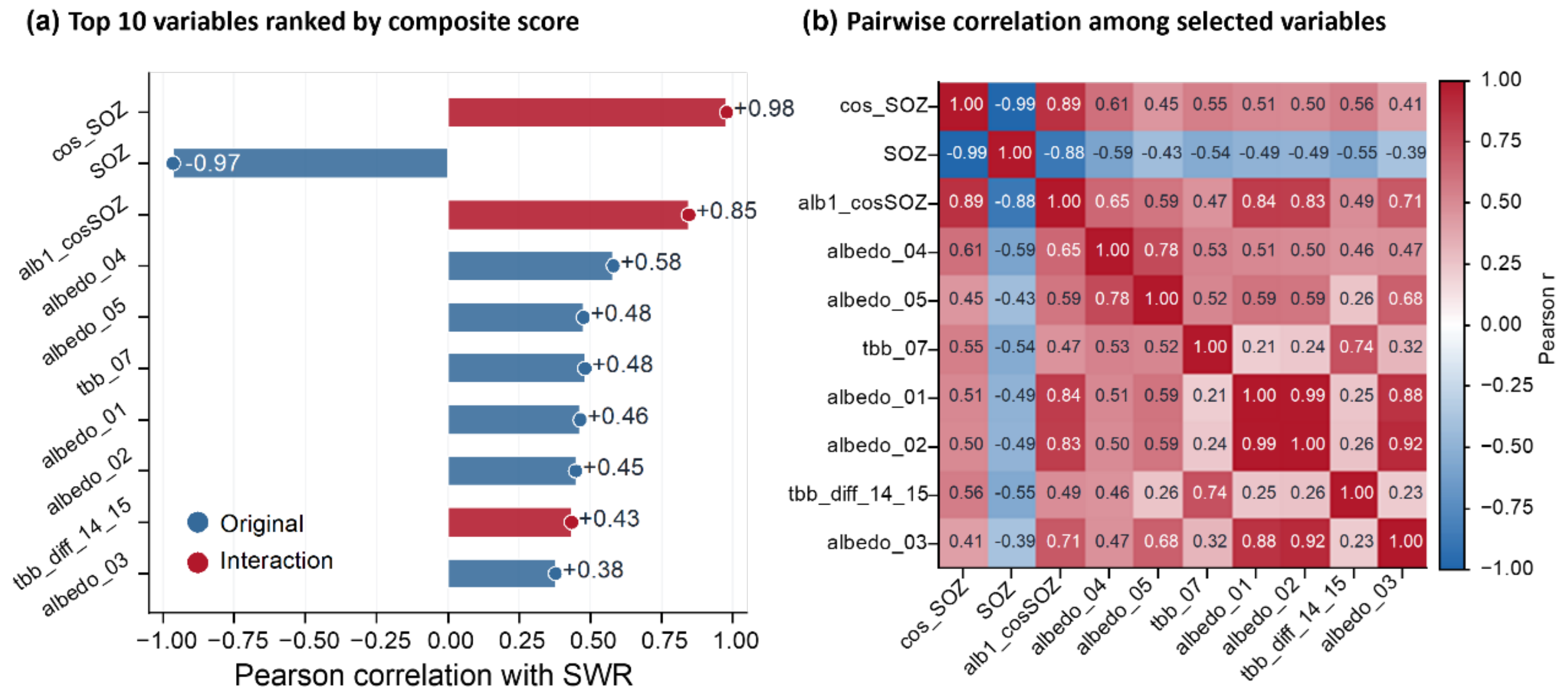

**Supplementary Fig. 1| Variable screening. a,** Pearson correlations for the ten highest-ranked candidate variables, ordered by the combined screening score. Blue and red denote raw and derived variables, respectively. **b,** Pairwise Pearson correlation matrix for the selected variables.

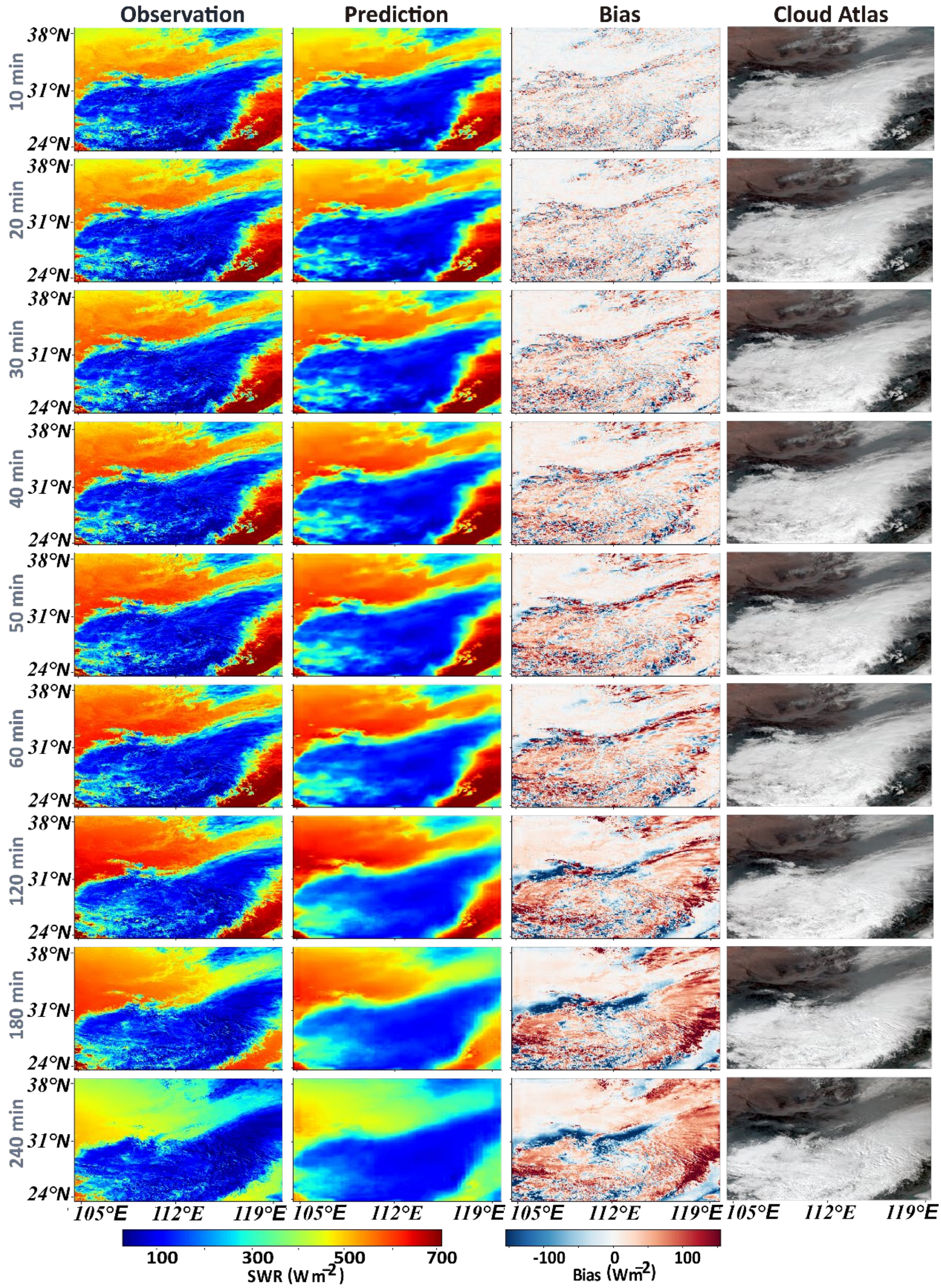

**Supplementary Fig. 2 | Exp1-TA forecast for the 2 January 2024 case.** Columns show the Level-2 SWR target, forecast, forecast-minus-target difference and cloud image; rows show selected lead times.

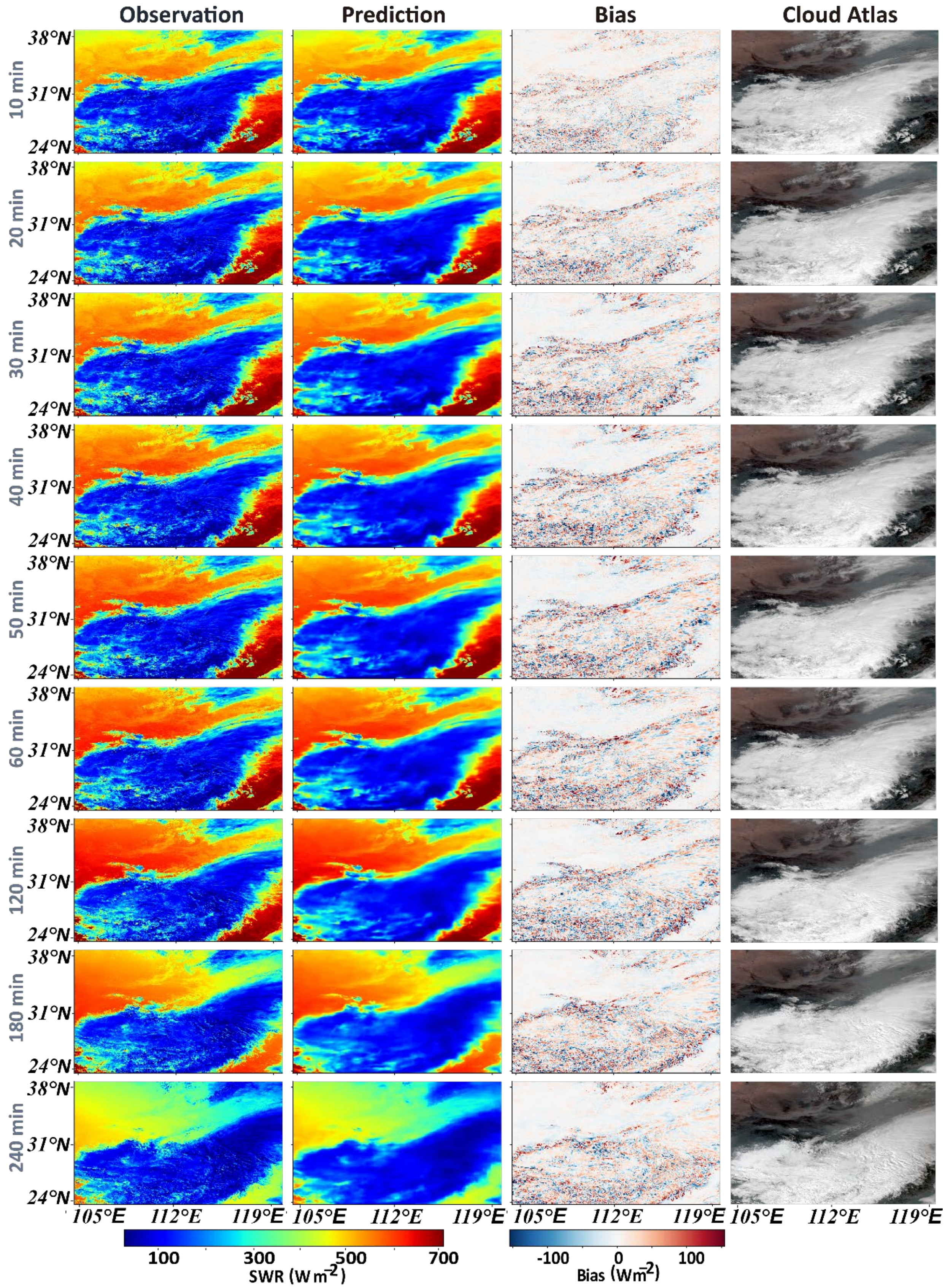

**Supplementary Fig. 3 | Exp2-TA forecast for the 2 January 2024 case.** Columns show the Level-2 SWR target, forecast, forecast-minus-target difference and cloud image; rows show selected lead times.